# ssVERDICT: Self-Supervised VERDICT-MRI for Enhanced Prostate Tumour Characterisation

Snigdha Sen[1]*, Saurabh Singh[2], Hayley Pye[3], Caroline M. Moore[3], Hayley Whitaker[3], Shonit Punwani[2], David Atkinson[2], Eleftheria Panagiotaki[1†], and Paddy J. Slator[1,4,5†]

1. Centre for Medical Image Computing, Department of Computer Science, University College London, London, United Kingdom
2. Centre for Medical Imaging, University College London, London, United Kingdom
3. Department of Targeted Intervention, Division of Surgery and Interventional Science, University College London, London, United Kingdom
4. Cardiff University Brain Research Imaging Centre, School of Psychology, Cardiff University, Cardiff, United Kingdom
5. School of Computer Science and Informatics, Cardiff University, Cardiff, United Kingdom
*Corresponding author: snigdha.sen.20@ucl.ac.uk
†These authors contributed equally to this work
Word count: 3143

## Abstract

**Purpose:** Demonstrating and assessing self-supervised machine learning fitting of the VERDICT (Vascular, Extracellular and Restricted DIffusion for Cytometry in Tumours) model for prostate.

**Methods:** We derive a self-supervised neural network for fitting VERDICT (ssVERDICT) that estimates parameter maps without training data. We compare the performance of ssVERDICT to two established baseline methods for fitting diffusion MRI models: conventional nonlinear least squares (NLLS) and supervised deep learning. We do this quantitatively on simulated data, by comparing the Pearson's correlation coefficient, mean-squared error (MSE), bias, and variance with respect to the simulated ground-truth. We also calculate *in vivo* parameter maps on a cohort of 20 prostate cancer patients and compare the methods' performance in discriminating benign from cancerous tissue via Wilcoxon's signed-rank test.

**Results:** In simulations, ssVERDICT outperforms the baseline methods (NLLS and supervised DL) in estimating all the parameters from the VERDICT prostate model in terms of Pearson's correlation coefficient, bias, and MSE. *In vivo*, ssVERDICT shows stronger lesion conspicuity across all parameter maps, and improves discrimination between benign and cancerous tissue over the baseline methods.

**Conclusion:** ssVERDICT significantly outperforms state-of-the-art methods for VERDICT model fitting, and shows for the first time, fitting of a complex three-compartment biophysical model with machine learning without the requirement of explicit training labels.

*Keywords--* Diffusion MRI, Microstructure Estimation, Self-Supervised Learning, Prostate Cancer

1. **Introduction**

Prostate (PCa) characterisation is reliant on invasive biopsy, but in recent years, multiparametric MRI (mp-MRI) has become established in the diagnostic pathway for localisation and staging of clinically-significant PCa (csPCa) [1]. Diffusion MRI (dMRI) is a powerful component of mp-MRI, measuring the motion of water molecules in biological tissues to infer information about local microstructure. Advanced multi-compartment models of the dMRI signal can estimate parameters relating to specific microstructural properties such as cell size, density and vasculature [2]. Such models enable non-invasive analysis of similar metrics to those typically only accessible by histology, and have been shown to reduce the need for invasive biopsies in breast [3] and prostate cancer [4].

Microstructural information can be extracted by designing models with parameters corresponding to these biophysically-relevant metrics, which are then estimated by fitting the models to dMRI data. These models tend to be nonlinear with many free parameters, and dense q-space sampling is required for accurate description of the microstructure, which involves time-consuming processes. This means that parameter estimation becomes a difficult inverse problem, scaling with both voxel number and model complexity. Additionally, parameter estimation requires an optimisation-based procedure, typically relying on nonlinear least squares (NLLS) curve fitting, which is computationally expensive and prone to estimation errors [5]. These challenges when obtaining and examining dMRI data hinder the clinical translation of these methods.

Recent work has utilised deep learning (DL) techniques to solve this parameter estimation inverse problem. These algorithms learn the mapping between the q-space data and the microstructural parameters of the dMRI model. Pioneering work on q-space learning by Golkov et al. [6] estimated model parameters using a multilayer perceptron (MLP), an approach that has since been widely used for ultrafast model fitting [7-9]. Convolutional neural networks (CNNs) have also been used with supervised learning for dMRI model fitting [10], as have transformers [11]. This approach has been used to fit both simple exponential models, as well complex biophysical models such as Neurite Orientation Dispersion and Density Imaging (NODDI) [12] and the Spherical Mean Technique (SMT) [13]. However, these methods are significantly affected by the underlying distribution of the training data, which can introduce biases in the parameter estimates [5,13].

Self-supervised learning techniques can circumvent this issue, as they do not rely on explicitly labelled training data, instead extracting labels from the input data itself. This approach has been successful for microstructural parameter estimation with the simple bi-exponential intravoxel incoherent motion (IVIM) model [5], [14-16]. However, despite these numerous IVIM examples, and in contrast to supervised model fitting, self-supervised model fitting has not been demonstrated for complex biophysical models.

Here we introduce a self-supervised approach to fit the Vascular, Extracellular and Restricted DIffusion for Cytometry in Tumours (VERDICT) model for the prostate, a complex three-compartment biophysical model [17]. We refer to our method as ssVERDICT. The VERDICT framework, currently in clinical trials for PCa, requires robust model fitting to estimate microstructural metrics such as cell size, intracellular volume fraction and diffusivity. These have previously been estimated via NLLS and supervised DL approaches [18,19], but the complexity of VERDICT increases its susceptibility to the aforementioned limitations of these techniques.

This is the first work demonstrating self-supervised fitting beyond simple exponential models. We show that ssVERDICT achieves higher accuracy and reduced bias when estimating microstructural parameters using ground truth simulations. On real data, ssVERDICT achieves discrimination of cancerous tissue from benign at a higher confidence level on a dataset of 20 PCa patients, highlighting the potential of the method for clinical translation.

## 2. **Methods**

In this section, we firstly introduce the VERDICT model for prostate, a three-compartment biophysical dMRI model. We then outline how the simulated data was generated, and how the patient data was acquired. We discuss the two baseline fitting methods (conventional NLLS fitting and supervised deep learning), followed by our novel self-supervised fitting method, ssVERDICT. Finally, we give details on the pre-processing steps, ROI selection and the evaluation metrics used.

*2.1 VERDICT Model*

The VERDICT prostate model is the sum of three parametric models, describing the dMRI signal as intracellular (IC), extracellular-extravascular (EES) and vascular (VASC) water [17]. The total signal is:

$$S = f_{VASC} S_{VASC}(d_{VASC}, b) + f_{IC} S_{IC}(d_{IC}, R, b, \Delta, \delta) + f_{EES} S_{EES}(d_{EES}, b) \quad (1)$$

where $f_i$ is the volume fraction and $S_i$ is the signal with no diffusion weighting from water molecules in population $i$, where $i =$ IC, VASC or EES. The vascular signal fraction, $f_{VASC}$, is computed as $1 - f_{IC} - f_{EES}$, since $\sum_{i=1}^{3} f_i = 1$ and $0 \leq f_i \leq 1$ [17]. Here $b$ is the b-value, $\Delta$ is the gradient pulse separation and $\delta$ is the gradient pulse duration.

The mathematical signal forms are as follows:

$$S_{VASC} = \frac{\sqrt{\pi}}{2} \cdot \frac{\phi\left(\sqrt{b\left(-\left(\Delta - \delta/3\right)(\gamma\delta)^2 d_{VASC}\right)}\right)}{\sqrt{b\left(-\left(\Delta - \delta/3\right)(\gamma\delta)^2 d_{VASC}\right)}} \quad (2)$$

$$S_{IC} = \exp\left(-\frac{2\gamma^2 G^2}{d_{IC}} \sum_{m=1}^{\infty} \frac{\alpha_m^{-4}}{\alpha_m^2 R^2 - 2}\left[2\delta - \frac{2 + e^{-\alpha_m^2 d_{IC}(\Delta-\delta)} - 2e^{-\alpha_m^2 d_{IC}\delta} - 2e^{-\alpha_m^2 d_{IC}\Delta} + 2e^{-\alpha_m^2 d_{IC}(\Delta+\delta)}}{\alpha_m^2 d_{IC}}\right]\right) \quad (3)$$

$$S_{EES} = e^{-b d_{EES}} \quad (4)$$

where $\phi$ is the error function $\phi(z) = \int_0^z \exp(-t^2)\, dt$ and $\alpha_m^2$ is the $m^{th}$ root of $(\alpha R)^{-1} J_{\frac{3}{2}}(\alpha R) = J_{\frac{5}{2}}$, where $J_n(x)$ is a Bessel function of the first kind [2].

The spherical mean version of the VERDICT model represents the IC component as spheres of radius $R$ (using the GPD approximation [20]) with intra-sphere diffusivity fixed at $d_{IC} = 2\ \mu m^2/ms$. The EES component is represented as Gaussian isotropic diffusion with effective diffusivity $d_{EES}$, and the vascular component as spherically-averaged randomly oriented sticks with intra-stick diffusivity fixed at $d_{VASC} = 8\ \mu m^2/ms$ [21]. By fitting the model to dMRI data, we estimate four model parameters: $f_{EES}$, $f_{IC}$, $R$ and $d_{EES}$.

*2.2 Patient Data*

The study was performed with the approval of the local ethics committee embedded within the INNOVATE clinical trial (NCT02689271) [22], which included men suspected of having csPCa. For this study, we randomly selected 20 patients from the INNOVATE cohort with biopsy-confirmed csPCa. VERDICT-MRI was performed on a 3T MRI system (Achieva; Philips, Best, the Netherlands), using a pulsed-gradient spin echo sequence. The imaging parameters, as published in [4,22-24], were as follows: repetition time (TR), 2482–3945 ms; field of view, $200 \times 220$ mm; voxel size, $1.3 \times 1.3 \times 5$ mm; no interslice gap; acquisition matrix, $176 \times 176$. The optimised VERDICT acquisition protocol for prostate is: $b$, 90, 500, 1500, 2000 and 3000 s/mm$^2$; $\delta$, 3.9, 3.9, 11.4, 23.9, 14.4, 18.9 ms; $\Delta$, 23.8, 23.8, 31.3, 43.8, 34.3, 38.8 ms [23]. For each of the five combinations of $b/\delta/\Delta$ we used the minimum possible echo time (TE), giving TEs of 50–90 ms, and we acquired a separate $b = 0$ image for each TE, resulting in 10 image volumes.

*2.3 Simulated Data*

We generated synthetic datasets for quantitative analysis using the VERDICT model with added Rician noise. We first simulated datasets with SNR levels ranging from 10-100 so we could test the robustness of the methods to noise. We then set SNR = 50 for the final simulated dataset. We created 100,000 signals from uniform VERDICT parameter distributions within biophysically realistic parameter ranges: $f_{EES} = [0.01, 0.99]$, $f_{IC} = [0.01, 0.99]$, $R = [0.01, 15]\mu m$ and $d_{EES} = [0.5, 3]\mu m^2/ms$. We simulated dMRI data using the same acquisition protocol as the patient data [22]. The parameters were drawn from uniform (rather than *in vivo*) distributions to minimise bias in the resulting parameter estimates [5,13].

*2.4 Conventional Iterative Fitting*

We fit the VERDICT model via NLLS using custom code in MATLAB (The Mathworks Inc., Natick, Massachusetts, USA), with parameter constraints as given in Sec. 2.3 using the 'lsqcurvefit' function as in [17,19]. Prediction for the whole unmasked dMRI dataset (roughly $5 \times 10^5$ voxels) took ~140 s per subject (Apple M1 Pro).

*2.5 Supervised Deep Learning*

Supervised techniques approximate the function $f$ that maps the measurement **S** to its corresponding ground truth parameters, **x**, by minimising the difference between the ground truth parameter values (training labels) and the parameter estimates (network output). The training loss is calculated as the mean squared error (MSE) between the estimated and ground truth values. We use an MLP architecture, implemented in Python 3.7.13 using the 'MLPregressor' in scikit-learn 0.23, as in [18,24-26].

The input of the deep neural network (DNN) is a vector of dMRI signals for each combination of $b$, $\Delta$, $\delta$ (a total of 10 in this case), followed by three fully-connected hidden layers with 150 neurons [18,24-26], and a final regression layer with four output neurons (equal to the number of parameters to be estimated). The DNN is trained on 100,000 synthetic signals (split into 80% for training and 20% for validation), with values for the model parameters randomly chosen from the ranges given in Sec. 2.3. We performed the optimisation with the ADAM method for 1000 epochs (adaptive learning rate with initial value of 0.001; one update per minibatch of 100 voxels; early stopping to mitigate overfitting; and momentum = 0.9). For the final parameter computation, we used the DNN at the epoch with minimum validation loss. The creation of the training set and training of the DNN (which was performed only once) took ~200 s. Prediction of the trained DNN for the whole unmasked dMRI dataset took ~30 s per subject.

*2.6 Self-Supervised Deep Learning*

Self-supervised methods compute $f$ by minimising the difference between the noisy MR signals (network inputs) and noise-free signal estimates reconstructed from the estimated parameters (network outputs). The training loss is equivalent to the MSE between the predicted signal, $\hat{S}$, and the input signal $S$ [14]. Here, network training and inference is performed on the same dataset, mimicking the NLLS approach.

We implemented a fully-connected neural network with three hidden layers, each with 10 neurons (equal to the number of image volumes), using PyTorch 1.12.1. The output layer is fed into the VERDICT model equation to generate the predicted signal $\hat{S}$. Crucially, this requires coding the VERDICT model in a differentiable form to enable backpropagation. For this, we formulate the intricate signal equations for VERDICT's 'sphere' and 'astrosticks' compartments (Eqs. 2,3) as PyTorch tensor functions, so that multi-dimensional tensors of batched parameter values can be inputted to yield output tensors of batched predicted signals. A schematic of ssVERDICT is given in Fig. 1.

For the final parameter estimation, we used the normalised input data, the ADAM optimiser and the DNN at the epoch with minimum validation loss. We optimised the DNN by backpropagating the MSE between $S$ and $\hat{S}$, where $\hat{S}$ is reconstructed via the VERDICT model from the parameter estimates. We chose a learning rate of 0.0001 and the network was trained until 10 consecutive epochs occurred without any improvement in loss, before terminating to prevent overtraining. We used dropout ($p = 0.5$) to prevent overfitting and constrained the parameter values to the ranges in Sec. 2.3 using the PyTorch clamp function. Training and prediction for the whole unmasked dMRI dataset took ~50 s per subject.

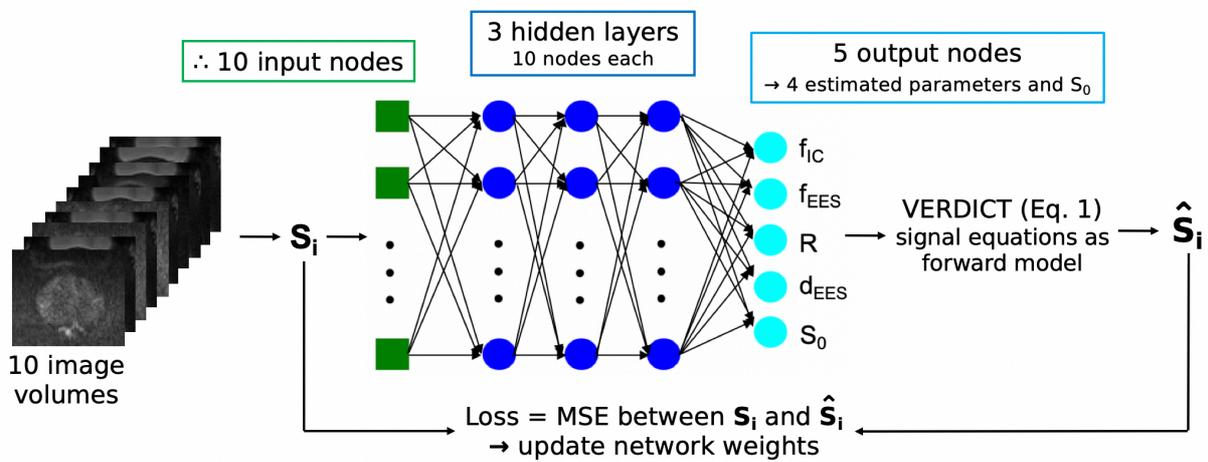

Figure 1: Schematic of our self-supervised network. The input to the neural network is the signal extracted from 10 signal volumes, therefore there are 10 input notes. The network has three hidden layers, each with 10 nodes. The final layer has five nodes, corresponding to the four estimated VERDICT parameters and $S_0$, the signal with no diffusion weighting. To reconstruct the signal ($\hat{S}$), the complex VERDICT signal equations (Eqs 1-4) are written in differentiable form so that it can be incorporated as a layer

*2.7 Data Pre-processing*

The pre-processing pipeline, as published in [4,22-24], included denoising of the raw DW- MRI using MP-PCA [27] as implemented within MrTrix3 [28] 'dwidenoise', and then correction for Gibbs ringing [29] with custom code in MATLAB. To reduce potential artefacts caused by patient movement during scanning and eddy current distortions, we applied mutual-information rigid and affine registration using custom code in MATLAB [30]. We normalised the data by dividing the dMRI volumes by their matched $b$ = 0. As we use the spherical mean version of the VERDICT model, we spherically averaged the data to produce 10 image volumes, where each volume was a 3D image consisting of 14 slices.

*2.8 ROIs*

Patients were biopsied based on their mp-MRI score as reported by two board-certified experienced uroradiologists (reporting more than 2,000 prostate MR scans per year). The regions of interest (ROI) were drawn by a board-certified study radiologist (S. Singh) using a pictorial report made by the clinical uroradiologist, and confirmed as cancerous retrospectively via targeted biopsy. An additional ROI was located for each patient in an area of benign tissue to be used for comparison, after a review of the sampling biopsy result confirmed the absence of tumour on the contralateral side.

*2.9 Evaluation Metrics*

We quantitatively compared the performance of the three parameter estimation methods via a variety of evaluation metrics: (1) Pearson's correlation coefficient (2) MSE (3) bias and (4) variance, all with respect to ground truth parameter values used for the simulated data. The formulae for the metrics used are given below:

$$\text{MSE} = \frac{1}{N}\sum_{i=0}^{N}(O_i - E_i)^2 \qquad (5)$$

$$\text{Bias} = \frac{1}{N}\sum_{i=0}^{N}(O_i - E_i) \qquad (6)$$

$$\text{Variance} = \frac{1}{N}\sum_{i=0}^{N}(O_i - \bar{O})^2 \qquad (7)$$

where $O$ is the ground truth parameter value, $E$ is the estimated value, and $N$ is the number of samples.

We performed discrimination between tissue types *in vivo* using the Wilcoxon's signed-rank test (preceded by the Shapiro-Wilk test for normality).

## 3. **Results**

Figure 2 shows estimated VERDICT parameters via each fitting method plotted against randomly-generated ground truth parameter values (Sec. 2.3). The Pearson's correlation coefficients $r$ are highest for all four VERDICT parameters when fitted via ssVERDICT. We also observe higher $r$ values for supervised DL fitting over NLLS.

Note the horizontal lines in the NLLS $R$ correlation plot, corresponding to the $R$ values in the grid-search stage of the NLLS fitting.

Table 1 gives the MSE, bias and variance values for all four fitted parameters, with each of the fitting methods. We observe lower bias and MSE across all parameters via ssVERDICT, and lower variance in estimating $f_{EES}$ and $d_{EES}$. However, supervised DL fitting achieves lowest variance in estimating $f_{IC}$ and $R$.

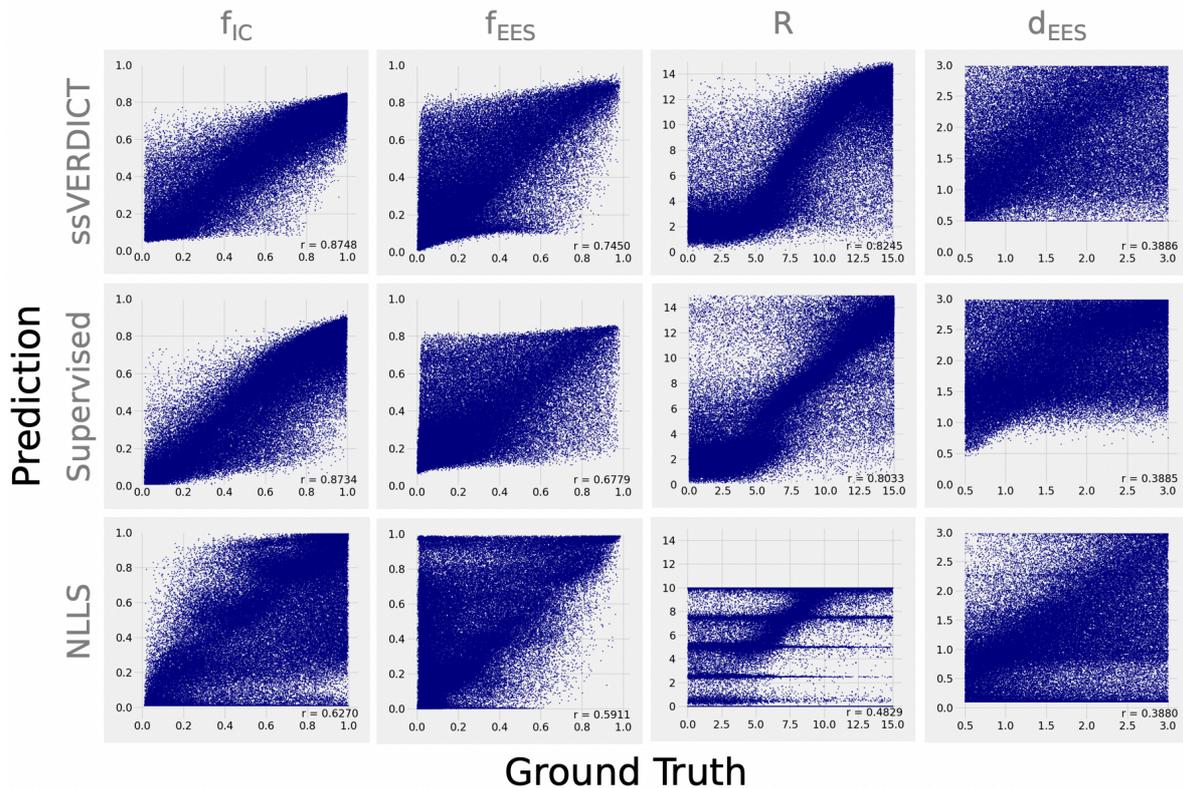

Figure 2: Scatterplot of simulated ground truth parameter values and predicted values via the three fitting methods. We observe higher Pearson's correlation coefficient r when using ssVERDICT for all four estimated parameters.

Figure 3 shows *in vivo* maps of the four fitted VERDICT parameters and calculated $f_{VASC}$. ssVERDICT shows strong lesion conspicuity for the $f_{IC}$ and $f_{EES}$ maps, and reasonable conspicuity on the $f_{VASC}$, $R$ and $d_{EES}$ maps. The supervised DL method achieves strong lesion conspicuity for $f_{EES}$ and $f_{VASC}$, and the NLLS method for $f_{IC}$ and $f_{EES}$.

Figure 4 shows boxplots of the fitted VERDICT parameters, in benign and cancerous prostate tissue for a dataset of 20 patients. All three methods can discriminate between tissue types to a high level of significance with $f_{IC}$ and $f_{EES}$. ssVERDICT increases discrimination between benign and cancerous prostate tissue when compared to NLLS and supervised fitting in two ways:

1. Achieves statistical significance at $p < 0.001$ with extracellular-extravascular diffusivity ($d_{EES}$), which is not shown by the other techniques
2. Shows statistically significant differences at $p < 0.05$ for cell radius ($R$), which is not seen with supervised DL or NLLS fitting.

Table I: MSE, bias and variance values calculated between simulated ground truth and predictions obtained via each fitting method. We find ssVERDICT achieves the lowest MSE and bias across all four parameters, and lowest variance for $f_{EES}$ and $d_{EES}$ also.

| MSE | | | | |
|---|---|---|---|---|
| **Method** | $f_{IC}$ | $f_{EES}$ | $R$ | $d_{EES}$ |
| NLLS | 0.1232 | 0.1137 | 17.6976 | 0.9905 |
| Supervised DL | 0.0714 | 0.0994 | 6.8860 | 0.7489 |
| ssVERDICT | **0.0289** | **0.0362** | **5.5278** | **0.7160** |
| Bias | | | | |
| **Method** | $f_{IC}$ | $f_{EES}$ | $R$ | $d_{EES}$ |
| NLLS | -0.0742 | 0.0680 | -0.9442 | -0.3624 |
| Supervised DL | -0.1008 | 0.1571 | -0.7152 | 0.3994 |
| ssVERDICT | **-0.0070** | **-0.0162** | **0.4002** | **0.2522** |
| Variance | | | | |
| **Method** | $f_{IC}$ | $f_{EES}$ | $R$ | $d_{EES}$ |
| NLLS | 0.0958 | 0.0942 | 17.5022 | 0.8649 |
| Supervised DL | **0.0459** | 0.0627 | **13.3562** | 0.6388 |
| ssVERDICT | 0.0655 | **0.0542** | 16.7244 | **0.4378** |

Figure 5 shows boxplots of the difference between the fitted VERDICT parameters and the ground truth values via the three fitting methods at varying SNR. We generally observe that across the VERDICT parameters, ssVERDICT results in estimates with a median difference closest to zero and smaller interquartile ranges. This suggests more accurate estimation via ssVERDICT across a range of SNR values. This also supports our decision to simulate data with an SNR of 50, as we show that parameter estimation remains robust across a wide range of SNRs.

## 4. Discussion

PCa diagnosis can be significantly improved by the introduction of non-invasive biomarkers derived from quantitative diffusion MRI [15,17,31]. However, clinical adoption of such techniques requires robust model fitting to avoid misdiagnosis [13,32]. This study presents a novel self-supervised fitting strategy (ssVERDICT) that can support biophysical multi-compartment dMRI models, demonstrated with the three-compartment VERDICT prostate [17,19]. Previously, self-supervised model fitting was limited only to simple exponential and biexponential models [5,14]. This is likely due to the difficulty involved in formulating complex signal equations (typical of biophysical models) as a differentiable forward model. Our work is a key step-change for self-supervised machine learning for dMRI model fitting, moving from simple models to complex multi-compartment biophysical models.

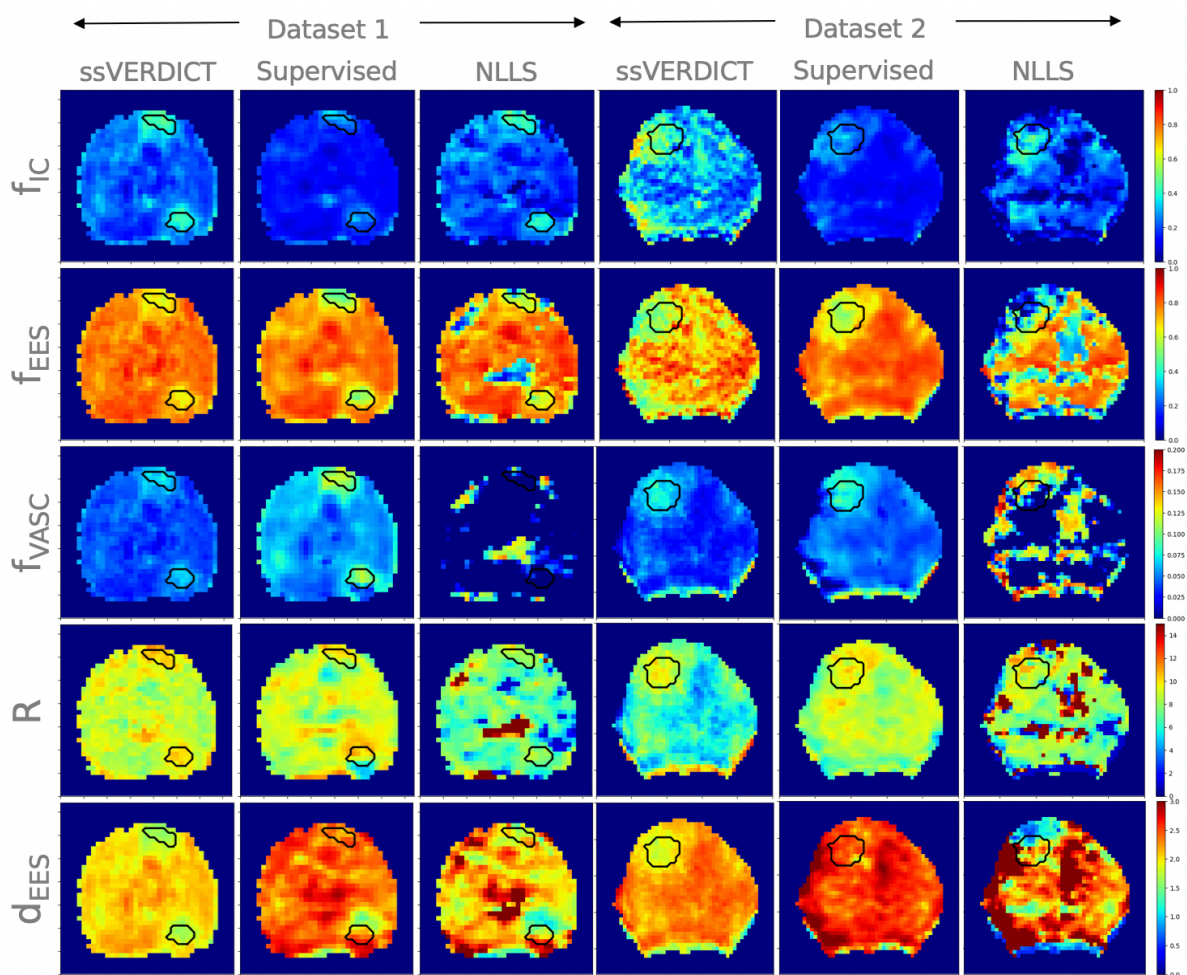

Figure 3: Parameter maps of the four fitted VERDICT parameters and calculated $f_{VASC}$ for two patients – dataset 1 shows a Gleason 3+3 grade tumour in the left anterior and 3+4 grade tumour in the right posterior peripheral zone, and dataset 2 shows a Gleason 4+3 grade tumour in the right posterior peripheral zone, and dataset 2 shows a Gleason 4+3 grade tumour in the right peripheral zone. We observe improved lesion conspicuity overall when using ssVERDICT, whilst supervised DL only shows strong tumour conspicuity for $f_{EES}$ and $f_{VASC}$, and NLLS only for $f_{IC}$ and $f_{EES}$.

We demonstrate that ssVERDICT outperforms the two gold-standard approaches for VERDICT model fitting (conventional iterative fitting (NLLS) and supervised DL fitting) [17-19,24-26,33] across a range of quantitative metrics. We also use ssVERDICT on clinical *in vivo* prostate data, showing excellent tissue discrimination between benign and cancerous tissue. Our work is the first to investigate model fitting estimation bias in prostate imaging, achieving reduced bias in comparison to supervised DL, as well as being faster than conventional NLLS fitting. PyTorch code for the complex VERDICT prostate model, as well as instructions on how to implement self-supervised fitting of other VERDICT-based biophysical models [18,33] is available at https://github.com/snigdha-sen/ssVERDICT. The differentiable form of the compartments can also be used to enable self-supervised fitting of other complex diffusion models, e.g. the 'sphere' for NODDI [34] and 'astrosticks' for the Soma And Neurite Density Imaging (SANDI) model [35].

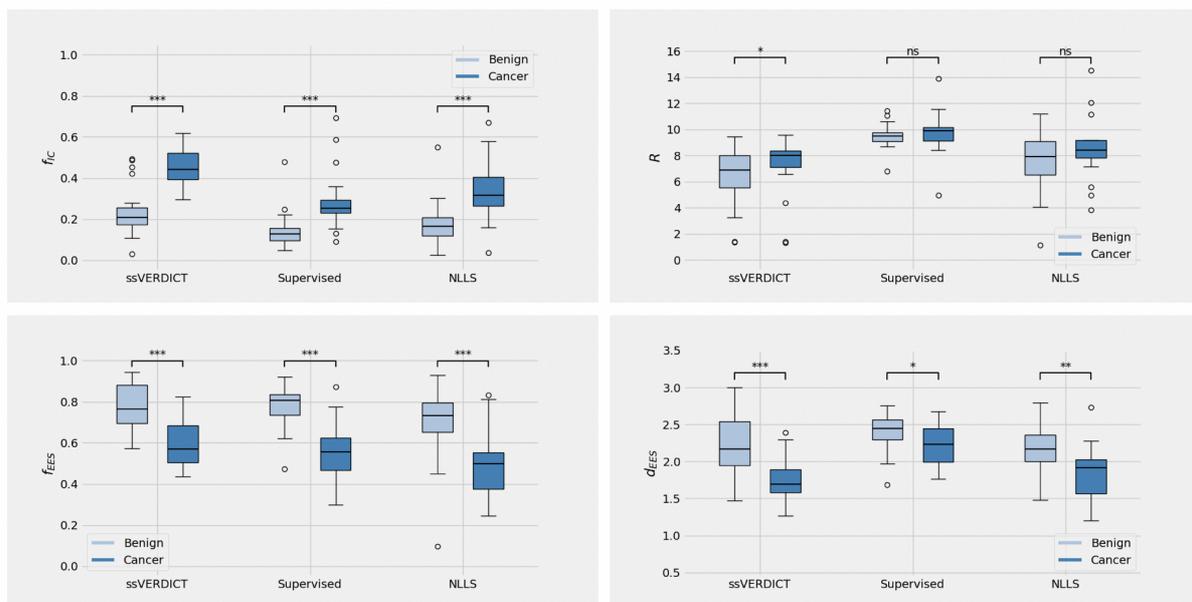

Figure 4: Boxplots of four fitted VERDICT parameter values in benign and cancerous tissue regions in a dataset of 20 PCa patients, calculated via the three fitting methods. We find that ssVERDICT maintains the high level of statistical significance achieved by the baseline methods when using $f_{IC}$ and $f_{EES}$ for tissue discrimination. ssVERDICT also improves the level of statistical significance with $d_{EES}$, and achieves statistical significance with $R$.

In simulations, ssVERDICT showed stronger correlations between estimated parameters and ground-truth values than the other two methods for all VERDICT parameters (Fig. 2). This suggests ssVERDICT can estimate the underlying microstructure more accurately than supervised DL and NLLS. We also found reduced bias and MSE across all four fitted VERDICT parameters when using ssVERDICT in comparison to the other methods, as well as lower variance when estimating $f_{EES}$ and $d_{EES}$ (Table 1). These results are in agreement with [5], [14] who found that self-supervised fitting of the simple IVIM model resulted in more accurate estimation than NLLS and lower bias than supervised DL. Our results demonstrate that this improvement in estimation translates to a significantly more complex multi-compartment model.

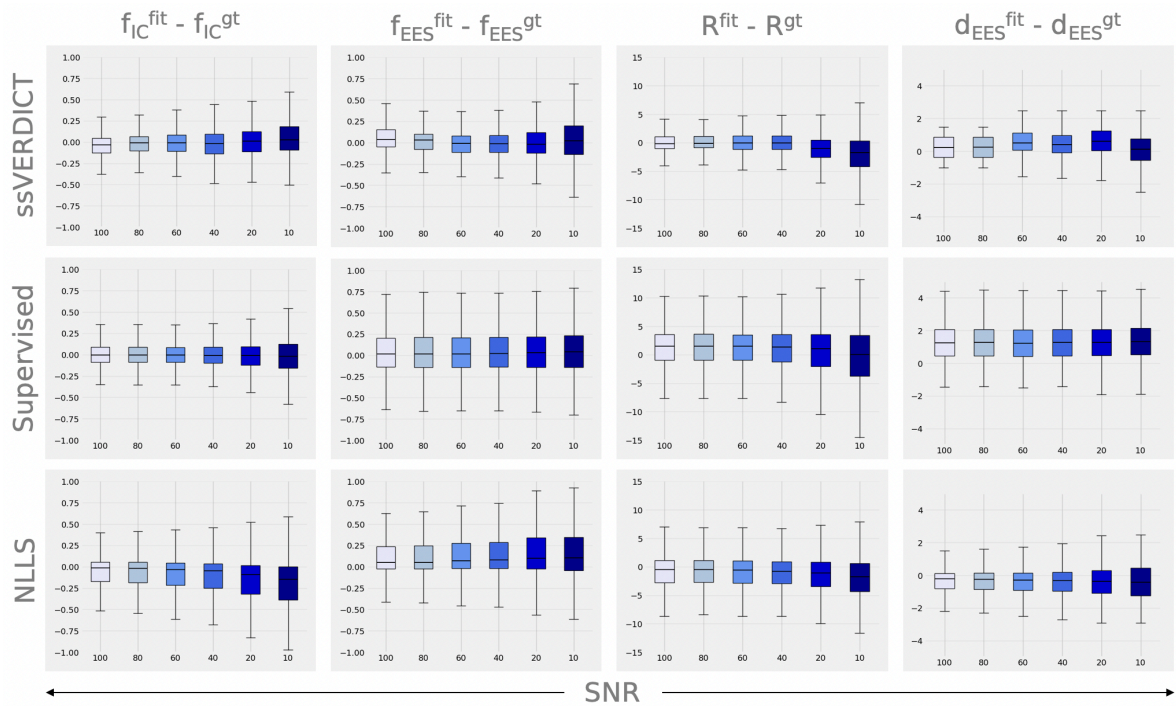

Figure 5: Boxplots of difference between fitted VERDICT parameter values and simulated ground truth using the three fitting strategies. We find median differences closest to zero and smaller interquartile ranges in general across the four parameters when using s ssVERDICT, suggesting more accurate fitting by our method across a range of SNR values.

Analysis of real patient data with ssVERDICT shows promising results *in vivo*, achieving the best tumour conspicuity over all VERDICT maps (e.g Fig. 3) and enhanced tissue type discrimination. We found higher statistical significance for $d_{EES}$ with ssVERDICT in comparison to other methods. For $f_{IC}$ and $f_{EES}$, ssVERDICT achieved statistical significance at $p < 0.001$, as did the baselines, and also achieved statistical significance for $R$. These results indicate that the improved accuracy in estimation with ssVERDICT in simulations transfers to patient data, demonstrating better discrimination between tissue types. This strongly suggests the benefits of our technique will translate to clinical practice, improving non-invasive tumour characterisation and hence further reducing invasive biopsies.

This work is limited primarily by the small size of the patient dataset, and the range of prostatic disease included. We also only focus on voxelwise methods, rather than extending to architectures that learn spatial correspondences in images such as CNNs or spatial transformers. Whilst a self-supervised CNN has been demonstrated for the IVIM model [15], [16] and supervised CNN methods have been widely used for dMRI model fitting, we instead focus on voxelwise fitting methods to enable a clear comparison between ssVERDICT and the currently used VERDICT fitting techniques in a controlled environment.

Future work will aim to increase statistical significance with a larger patient cohort, and incorporate a wider range of prostatic diseases [25], to test ssVERDICT's ability to accurately characterise tissue microstructure and maximise its potential clinical impact. We will also investigate fitting more complex biophysical models such as [18], [33] via a self-supervised CNN approach similar to [15], [17], to investigate potential further gains in fitting speed and accuracy.

In conclusion, our work shows that self-supervised fitting of the VERDICT prostate model performs better in simulations and *in vivo* data than baseline methods. This study is the first to extend self-supervised model fitting beyond highly simple models. Our results demonstrate that ssVERDICT provides accurate characterization of prostate tumour microstructure, contributing towards the ultimate goal of reducing the number of biopsies and improving patient care.

### Acknowledgement

This work was supported by the EPSRC-funded UCL Centre for Doctoral Training in Intelligent, Integrated Imaging in Healthcare (i4health) (EP/S021930/1) and the Department of Health's NIHR-funded Biomedical Research Centre at University College London Hospitals. This work is also funded by EPSRC, grant numbers EP/N021967/1, EP/R006032/1, EP/V034537/1; and by Prostate Cancer UK, Targeted Call 2014, Translational Research St.2, grant number PG14-018-TR2.

**Table 1:** MSE, bias and variance values calculated between simulated ground truth and predictions obtained via each fitting method. We find ssVERDICT achieves the lowest MSE and bias across all four parameters, and lowest variance for $f_{EES}$ and $d_{EES}$ also.

**Figure 1:** Schematic of our self-supervised network. The input to the neural network is the signal extracted from 10 signal volumes, therefore there are 10 input notes. The network has three hidden layers, each with 10 nodes. The final layer has five nodes, corresponding to the four estimated VERDICT parameters and $S_0$, the signal with no diffusion weighting. To reconstruct the signal ($\hat{S}$), the complex VERDICT signal equations (Eqs 1-4) are written in differentiable form so that it can be incorporated as a layer in the network, such that batches of signals can be inputted and batches of parameters outputted.

**Figure 2:** Scatterplot of simulated ground truth parameter values and predicted values via the three fitting methods. We observe higher Pearson's correlation coefficient r when using ssVERDICT for all four estimated parameters.

**Figure 3:** Parameter maps of the four fitted VERDICT parameters and calculated $f_{VASC}$ for two patients – dataset 1 shows a Gleason 3+3 grade tumour in the left anterior and 3+4 grade tumour in the right posterior peripheral zone, and dataset 2 shows a Gleason 4+3 grade tumour in the right posterior peripheral zone, and dataset 2 shows a Gleason 4+3 grade tumour in the right peripheral zone. We observe improved lesion conspicuity overall when using ssVERDICT, whilst supervised DL only shows strong tumour conspicuity for $f_{EES}$ and $f_{VASC}$, and NLLS only for $f_{IC}$ and $f_{EES}$.

**Figure 4:** Boxplots of four fitted VERDICT parameter values in benign and cancerous tissue regions in a dataset of 20 PCa patients, calculated via the three fitting methods. We find that ssVERDICT maintains the high level of statistical significance achieved by the baseline methods when using $f_{IC}$ and $f_{EES}$ for tissue discrimination. ssVERDICT also improves the level of statistical significance with $d_{EES}$, and achieves statistical significance with $R$.

**Figure 5:** Boxplots of difference between fitted VERDICT parameter values and simulated ground truth using the three fitting strategies. We find median differences closest to zero and smaller interquartile ranges in general across the four parameters when using s ssVERDICT, suggesting more accurate fitting by our method across a range of SNR values.